\documentclass[aps,pra,amsfonts,superscriptaddress,nofootinbib,preprint,letterpaper,showpacs,linenumbers]{revtex4-1}

\usepackage{xspace}
\usepackage{graphicx}
\usepackage{dcolumn}
\usepackage{srcltx}

\newcommand{\figab}{Fig.~}
\newcommand{\tabab}{Table~}
\newcommand{\secab}{Sec.~}
\newcommand{\eqab}{Eq.~}

\newcommand{\cfclb}{$\mathbf{CF_3Cl}$\xspace}
\newcommand{\cfcl}{$\mathrm{CF_3Cl}$\xspace}
\newcommand{\cfcla}{$\mathrm{CF_3Cl^-}$\xspace}

\newcommand{\cft}{$\mathrm{CF_3}$\xspace}

\newcommand{\dif}{\mathrm{d}}
\newcommand{\mi}{\mathrm{i}}

\newcommand{\rmat}{\textsl{R}-matrix\xspace}

\newcommand{\abini}{\textit{ab~initio}\xspace}

\newcommand{\bohr}{\thinspace a$_0$\xspace}

\newcommand{\eqref}[1]{(\ref{#1})}

\begin{document}
  \title{Dissociative electron attachment and vibrational excitation of \cfclb: Effect of two vibrational modes revisited}
	\author{\firstname{Michal} \surname{Tarana}}
  \email{michal.tarana@jila.colorado.edu}
	\affiliation{Department of Physics \& Astronomy, University of Nebraska, Lincoln, NE 68588, USA}
	\affiliation{JILA, University of Colorado and NIST, Boulder, Colorado 80309-0440, USA}

	\author{\firstname{Karel} \surname{Houfek}}
	\affiliation{Institute of Theoretical Physics, Faculty of Mathematics and Physics, Charles University in Prague, V Hole\v{s}ovi\v{c}k\'ach 2, Prague, Czech Republic}
	
	\author{\firstname{Ji\v{r}\'i} \surname{Hor\'a\v{c}ek}}
	\affiliation{Institute of Theoretical Physics, Faculty of Mathematics and Physics, Charles University in Prague, V Hole\v{s}ovi\v{c}k\'ach 2, Prague, Czech Republic}
	
	\author{\firstname{Ilya I.} \surname{Fabrikant}}
	\email{iif@unlserve.unl.edu}
	\affiliation{Department of Physics \& Astronomy, University of Nebraska, Lincoln, NE 68588, USA}
	\affiliation{Department of Physics and Astronomy, The Open University, Walton Hall, Milton Keynes MK7 6AA, UK }
	
	\begin{abstract}
		We present a study of dissociative electron attachment and vibrational excitation processes in electron collisions with the \cfcl molecule. The calculations are based on the two-dimensional nuclear dynamics including the C-Cl symmetric stretch coordinate and the \cft symmetric deformation (umbrella) coordinate. The complex potential energy surfaces are calculated using the \abini \rmat method. The results for dissociative attachment and vibrational excitation of the umbrella mode agree quite well with experiment while the cross section for excitation of the C-Cl symmetric stretch vibrations is about a factor of three low as compared to experimental data.
	\end{abstract}
	
	\maketitle
	
	\section{Introduction}
		Dissociative electron attachment (DEA) to polyatomic molecules typically involves
multidimensional nuclear dynamics. However, because of big computational work
necessary to obtain multidimensional complex ({\it i.e.}, including both real
and imaginary parts) energy surfaces, most of theoretical DEA calculations  
were performed in one-dimensional approximation. In these calculations 
it is usually
assumed that the DEA process involves one reaction (dissociating) coordinate,
roughly corresponding to one of the normal modes of the target molecule. 
This approximation is sometimes too crude, and sometimes completely unjustified.
Therefore, a lot of effort was devoted recently to calculations of 
multidimensional DEA dynamics 
\cite{tarana-cf3cl,kazansky-co2,haxton-h2o,chourou-acet,Cho09-hcn,Cho09-acet}.
These calculations address two important problems in the physics of DEA 
processes. First, we want to know which dissociation channels are the most
important and what is the energy range where a particular bond breaking 
can occur. This information is especially important for chemical control.
Second, we want to know the importance of different vibrational modes in a
particular DEA process and the final-state vibrational energy distribution
in the fragments. 

The most common method for studies of dynamics on multidimensional surfaces
is the wave-packet propagation technique \cite{McC83}. Recently, this
approach was used to calculate the DEA cross sections for several
polyatomic molecules, e.g. for CO$_2$ \cite{riomm2002} or H$_2$O
\cite{hzmrm2004}, employing the multi-configuration time-dependent
hartree (MCTDH) method \cite{mmc1990,bjwm2000}.
An alternative is to develop classical and semiclassical methods
for treatment of nuclear dynamics. We recently reformulated~\cite{tarana-cf3cl} 
the quantum method for DEA in terms of the time-independent   
Schr\"{o}dinger equation and connected this treatment with the classical 
approximation. The method was applied to the process
 \begin{equation}
   \mathrm{e+CF_3Cl}(\nu_2,\nu_3)\to\mathrm{CF_3Cl^-}\to\mathrm{CF_3}(\nu_2')+\mathrm{Cl^-}.
 \end{equation}
 Here $\nu_2$ and $\nu_3$ stand for the symmetric deformation vibrations 
(so-called ``umbrella'' mode) and the symmetric stretch vibrations, $\nu_2'$ 
represents the umbrella mode of the free \cft radical. The two-mode approximation for this process
can be justified by existing experimental data \cite{mann-cf3cl} on vibrational
excitation (VE) of this molecule.
The two-dimensional potential energy surface (PES) was calculated \abini.
However, we used a model semiempirical width (the imaginary part of the complex
PES). This led to some inconsistencies and instabilities in our calculations
discussed below. In the present paper we employ the \abini molecular
\rmat method for calculation of the complex PES. This allows us to remove
the deficiencies in our previous calculations and improve agreement with 
experimental data. We also calculate VE cross sections
for the C--Cl symmetric stretch and the umbrella mode.

The rest of the paper is organized as follows. In \secab\ref{sec:theo} we discuss construction of the complex PES from the \abini calculations and the  our theoretical 
approach to treatment of nuclear motion. In \secab\ref{sec:rmat} we discuss details of
our \rmat scattering calculations in the fixed-nuclei approximation.
In Sec. IV we present our results for DEA and VE.

	\section{Theoretical Approach}
	\label{sec:theo}
	In our previous work~\cite{tarana-cf3cl} we constructed a two-dimensional local complex potential (LCP) model for DEA and resonant VE of \cfcl. The model takes into account the C--Cl stretching vibrational mode and umbrella vibrational mode of the \cft radical. The LCP model~\cite{tarana-cf3cl} was based on a complex PES constructed from the one-dimensional potential curves using an arbitrary extension in the coordinate corresponding to the umbrella vibrations of the \cft fragment. In addition, the real part of the complex potential curve for 
the temporal anionic complex was obtained using the \abini methods for 
bound-states calculations, while the imaginary part was obtained by 
fitting~\cite{wilde} the experimental results of~\citet{mann-cf3cl}.
Cross sections of the DEA calculated using the two-dimensional model are factor of three higher than experimental values and one-dimensional non-local calculations~\cite{wilde}. This discrepancy was attributed to the inconsistencies between the real and imaginary part of the complex PES used in the model~\cite{tarana-cf3cl}. It is the aim of this work to construct a two-dimensional model of the nuclear dynamics along the same lines as in~\cite{tarana-cf3cl}, however using a more consistent complex PES.
	We performed 
molecular \abini \rmat scattering calculations in the fixed-nuclei 
approximation for a set of nuclear geometries including both degrees of freedom 
corresponding to the C--Cl distance $R$ and F--C--Cl angle $\vartheta$. We 
fitted the eigenphases obtained from these \rmat calculations at energies 
close to the resonance position to the Breit-Wigner formula with an
energy-dependent background~\cite{lane-thomas} and using the resonance 
position and width we constructed the complex PES. This approach is free of 
any presumption on the dependence of the complex PES on the F--C--Cl angle and removes the need for any arbitrary extension in this coordinate used in~\cite{tarana-cf3cl}.
In addition, the fixed-nuclei resonance width is calculated at the same level 
of the theory as the position.
	
	The \abini molecular \rmat method is well known and widely used for fixed-nuclei calculations of electron collisions with small and medium-sized molecules. We refer the reader to a recent review article by ~\citet{tennyson-rev} and to references therein for description of the method and its technical implementation.
	
	The eigenphase sums calculated using the \rmat method for a set of nuclear 
geometries were fitted using the Breit-Wigner formula taking into account the 
dominant dipole component of the potential in the outer region. 
%Since the 
%dipole moment around the equilibrium geometry is rather small (about 
%0.5\thinspace D), we employed the one-pole approximation to the \rmat~\cite{lane-thomas}.
The Breit-Wigner formula is equivalent to the one-pole approximation to the \rmat~\cite{lane-thomas}. This approximation is based on the assumption that at the energies close enough to the resonance position the \rmat can be well approximated by the following expression:
	\begin{equation}
		R(E)=R_0+\frac{\gamma_\lambda^2}{E_\lambda-E},
		\label{eq:resrmat}
	\end{equation}
	where $R_0$ is the background \rmat including all the terms due to remaining poles of the \rmat, $E_\lambda$ is position of the pole closest to the resonance, $\gamma_\lambda$ is corresponding amplitude and $E$ is the scattering energy. To obtain the Breit-Wigner formuula from \eqab\eqref{eq:resrmat}, we assume that the term $R_0$ is a slowly varying function of energy as it is sum over all the other \rmat poles. The eigenphase sum $\delta$ may then be expressed~\cite{lane-thomas} as
	\begin{equation}
		\delta(E)=\tan^{-1}\left(\frac{\frac{1}{3}\Gamma_\lambda}{E_\lambda+\Delta_\lambda-E}\right)-\phi(E),
		\label{eq:bwrmat}
	\end{equation}
	where $\Gamma_\lambda$ is the resonance width and $\Delta_\lambda$ is 
the level shift (amount by which the resonance energy is shifted from the pole 
$E_\lambda$). The first term of \eqab\eqref{eq:bwrmat} describes the resonance 
contribution and the second $(\phi(E))$ the potential scattering contribution. The relations between $\Gamma_\lambda$, $\Delta_\lambda$, $\phi$ and general solutions of the Schr\"odinger equation on the \rmat boundary are given in Ref.~\cite{lane-thomas}. In order to fit the width and the position of the resonance using this model we first solve the Schr\"odinger equation with the dipole potential in the outer region. The corresponding dipole moment for every nuclear geometry is obtained from the \abini calculation of target properties as a part of the fixed-nuclei scattering calculations. Having the solutions on the \rmat boundary we can establish the relation between the background phase shift $\phi(E)$, width $\Gamma_\lambda$, level shift $\Delta_\lambda$ and quantities in \eqab\eqref{eq:resrmat}. This allows us to fit the model \rmat amplitude $\gamma_\lambda$, pole $E_\lambda$ and constant background \rmat $R_0$ using the non-linear least-squares technique to the \abini eigenphase sum. These directly determine the width $\Gamma_\lambda$ and resonance energy $E_\lambda+\Delta_\lambda$. This fitting allows for a construction of the complex PES $U(R,r)-\mathrm{i}\Gamma(R,r)/2$ in the region where the anionic state is metastable. In the region where the resonance turns into a bound state, the corresponding bound state energy was calculated as well as the potential energy curve $V(\infty,r)$ of the free \cft fragment. The complex PES was constructed using cubic splines in two dimensions. In order to study the final-state interaction on the anionic surface during the DEA process, it was necessary to have the bound anionic PES also for large C--Cl internuclear separations, where the \abini results are not available. It was constructed by the extrapolation of the \abini results as described in~\cite{tarana-cf3cl} to satisfy the condition $U(R\to\infty,r)\to V(\infty,r)$. The bound part of the anionic potential energy surface was extrapolated to match the \cft fragment potential curve asymptotically, as discussed in~\cite{tarana-cf3cl}.
	
	The LCP calculations of the DEA and VE presented here were performed in a similar way as described in Ref.~\cite{tarana-cf3cl}. The basic equation of the LCP theory reads
	\begin{equation}
		[T_\rho+T_r+U(\rho,r)-\mi\Gamma(\rho,r)/2-E]\chi_E(\rho,r)=V_{dk}(\rho,r)\zeta_i(\rho,r),
		\label{eq:lcpbas}
	\end{equation}
	where $\rho$ and $r$ are the reaction coordinates introduced to decouple the two-dimensional operator of the nuclear kinetic energy~\cite{tarana-cf3cl}. $U(\rho,r)-\mi\Gamma(\rho,r)/2$ is the complex PES of the temporal anion, 
			$V_{dk}(\rho,r)=\sqrt{\Gamma(\rho,r)/2\pi}$ is the amplitude for electron capture into the resonance state and $\zeta_i(\rho,r)$ is the vibrational wave function of the neutral molecule in the initial state. $T_\rho+T_r$ is the operator of nuclear kinetic energy corresponding to our two-dimensional model as discussed in~\cite{tarana-cf3cl}. In our previous work~\cite{tarana-cf3cl} the wave function $\chi_E(\rho, r)$ was expanded in the basis of vibrational states $\phi_\nu(r)$ of the \cft fragment in the harmonic approximation given by equation
	\begin{equation}
		[T_r+V^\mathrm{h}(\infty,r)-\epsilon_\nu]\phi_\nu(r)=0,
		\label{eq:fragvib}
	\end{equation}
	where $\epsilon_\nu$ are the corresponding eigenenergies
$\epsilon_\nu=D_e+\omega_2^f(\nu+1/2)$, $\omega_2^f$ is the harmonic frequency 
of the \cft radical umbrella mode and $V^\mathrm{h}(\infty,r)$ is the 
corresponding free \cft radical potential curve in the harmonic approximation 
with the minimum corresponding to the C--Cl bond dissociation energy. As it is explained in~\cite{tarana-cf3cl}, projection of \eqab\eqref{eq:lcpbas} on $\phi_\nu(r)$ then yields a set of coupled differential equations for channel wave functions of the variable $\rho$ with coupling potential $U_{\nu\nu'}(\rho)$ given by the equation
  \begin{equation}
    U_{\nu\nu'}(\rho)=\int\phi_\nu(r)\left[U(\rho,r)-\mi\Gamma(\rho,r)/2-V^\mathrm{h}(\infty,r)\right]\phi_{\nu'}(r)\dif r.
    \label{eq:chancoup}
  \end{equation}
  One note should be made at this point regarding the asymptotic behavior of the coupling potential. The PES in~\cite{tarana-cf3cl} was extrapolated in such way that $U(R\to\infty,r)\to V(\infty,r)$. The extrapolation asymptotically matches the \abini potential energy curve of the \cft fragment, not its harmonic approximation. As a consequence, $\lim_{\rho\to\infty}U_{\nu\nu'}(\rho)\neq 0$, as can be seen in~\eqab\eqref{eq:chancoup}. Since the coupling of different vibrational channels of the \cft radical doesn't vanish for $\rho\to\infty$, population of different vibrational states of \cft produced by the DEA process does not converge well. This deficiency is corrected in the present work. The harmonic approximation for the potential energy curve $V^{\text{h}}(\infty,r)$ of the \cft radical is not used and the full \abini potential curve $V(\infty, r)$ is employed in both \eqab\eqref{eq:fragvib} and \eqab\eqref{eq:chancoup}. Therefore, in our present calculations we expand the wave function $\chi_E(\rho,r)$ in the basis set of eigenfunctions $\varphi_\nu(r)$ given by the equation
  \begin{equation}
    [T_r+V(\infty,r)-\varepsilon_\nu]\varphi_\nu(r)=0,
  \end{equation}
  where $\varepsilon_\nu$ are the vibrational energies of the \cft fragment without the harmonic approximation. The corresponding coupling potential has the following form:
  \begin{equation}
    U_{\nu\nu'}(\rho)=\int\varphi_\nu(r)\left[U(\rho,r)-\mi\Gamma(\rho,r)/2-V(\infty,r)\right]\varphi_{\nu'}(r)\dif r.
  \end{equation}
  Therefore, in the present work the extrapolated PES is consistent with the asymptotic treatment of the \cft fragment that allows for convergence of the population of vibrational states of \cft produced by the DEA process.
  
	In the present work we also use a different method to solve the system of coupled radial equations. In Ref.~\cite{tarana-cf3cl} we used 
 direct outward integration of the system of differential equations from 
the inner region and inward integration from the asymptotic region with
subsequent matching of the solutions to satisfy the boundary conditions. 
The direct integration has a limitation in number of channels 
included in the calculation. With increasing number of closed channels 
included, their exponentially increasing contribution starts to be more 
pronounced and the calculation becomes unstable. 

Here we employ the multichannel version of the exterior complex scaling 
(ECS) method in the discrete variable representation (DVR) 
basis set~\cite{rescigno-ecs-dvr}, 
 which is free of this problem. This method has previously been 
successfully used in the context of nonlocal resonance model for calculations 
of DEA and VE in case of diatomic molecules~\cite{houfek-ecs}.
	
\section{Scattering calculations in the fixed-nuclei approximation}
\label{sec:rmat}
	In order to obtain the two-dimensional complex PES necessary to construct the LCP model, we performed the \rmat scattering calculations in the fixed nuclei approximation for a set of nuclear geometries important for the DEA and VE. For every nuclear geometry we calculated the eigenphase sum in the energy interval around the fixed-nuclei resonance and fitted it to the Breit--Wigner formula with background as described above. The resonance position and width as a function of the nuclear coordinates represent the complex PES used in the local complex approximation. In the region where the \cfcla anion is stable against autodetachment and the potential becomes real, the surface is represented by the bound-state energies calculated using the \rmat approach.

	As it was discussed previously~\cite{tarana-cf3cl}, in our calculations the PES $V$ and $U$ for the neutral molecule and for the anion are represented using two coordinates: C-Cl internuclear separation $R$ and the distance between the C atom and the plane formed by the fluorine atoms $r=-R_{CF}\cos\vartheta$, where $\vartheta$ is the F-C-Cl angle and $R_{CF}$ is the F-C bond length. Since we do not include the C-F stretching mode into our considerations, $R_{CF}$ is fixed and set to the value 1.342\thinspace\AA{} corresponding to the equilibrium geometry of the neutral~\cfcl. The fixed-nuclei \rmat calculations were performed for a two-dimensional region of nuclear coordinates with $R$ from 3\bohr to 12\bohr and with $\vartheta$ from $55^\circ$ to $90^\circ$. In the present work we consider excitation of the low vibrational states of the \cft fragment only, therefore we don't take into account any effects caused by the flipping of the radical ($\vartheta>90^\circ$).
	
	Calculations were performed using Gaussian type orbitals (GTOs) and the UK polyatomic \rmat code~\cite{tennyson-rev}. The highest symmetry available in the polyatomic code is $C_s$ which is an abelian subgroup of the true $C_{3v}$ symmetry of \cfcl.

	\subsection{Target representation}
	The \cfcl was represented using Hartree--Fock (HF) molecular orbitals (MOs). In order to construct a target model sufficient for the purpose of the dynamical calculations, we performed several tests with different target models to select the best compromise between the quality of the target representation and the computational tractability of the $(N+1)$ - particle problem.
	
	The \cfcl target states were represented using a complete active space 
(CAS) configuration interaction (CI) wave function. The \cfcl molecule contains
50 electrons. Only 18 of them belong to the inner shells, 32 remaining 
electrons form the valence shells and in principle can contribute to the 
chemical bonds. This complicates the construction of the target CI model in 
several aspects. Enough valence electrons should be included in the CI active 
space to treat the electron correlation properly. In addition, it is necessary 
to treat the target symmetrically and to ensure that all elements of the 
degenerated pairs of MOs are included in the CAS. On the other hand, inclusion 
of each orbital occupied in the HF ground state leads to a rapid increase in dimension of the target and anionic CI Hamiltonians. In addition, it is the aim of our fixed-nuclei calculations to study the dependence of the resonance on the nuclear geometry. This raises further the limitations on the $N+1$-particle CI calculation as we need to repeat it many times.
	
	The first CAS CI target model considered in our calculations includes eight active electrons which occupy four orbitals in the HF ground state determinant with the highest orbital energies. We allow these electrons to occupy five lowest virtual orbitals (VOs). Using the notation of the $C_s$ point group, this CAS model can be expressed as follows:
	\begin{equation}
	  (1\mathrm{a}'\dots 15\mathrm{a}')^{30}(1\mathrm{a}''\dots6\mathrm{a}'')^{12}(16\mathrm{a}'\dots21\mathrm{a}',7\mathrm{a}''\dots9\mathrm{a}'')^8
	\end{equation}
	or
		\begin{equation}
	  (1\mathrm{a}'\dots 15\mathrm{a}')^{30}(1\mathrm{a}''\dots6\mathrm{a}'')^{12}(16\mathrm{a}'\dots20\mathrm{a}',7\mathrm{a}''\dots10\mathrm{a}'')^8,
	\end{equation}
	as the ordering of VOs changes with nuclear geometry. Using the notation of the $C_{3v}$ point group, the active electrons are taken from the following set of HF orbitals: $(1a_2,10a_1,7e)$. The degenerated $e$-orbitals in this set are well localized on the chlorine atom, the $10a_1$-orbital is spread along the C--Cl bond and the $1a_2$ orbital contributes to the C--F bonds. The calculations were performed using the 6-311G* GTO basis set~\cite{six-basis, six-basis2}.

	Representation of the target in the subsequent scattering calculation can be qualified by comparison of several properties calculated using our model with the previously published results. To check our target representation, we calculated the vertical excitation energy (as the excited target states are used in the scattering calculations), dipole moment of the target ground state (as it represents the major contribution to the interaction with the projectile in the outer region) and static dipole polarizability which allows us to estimate the representation of the polarization effects in the scattering calculation. Comparison of values calculated using our models (at equilibrium nuclear geometry) with the previously published data is shown in \tabab~\ref{tab:targcompar}.
	\begin{table}[htb]
		\caption{Target properties calculated using different CAS CI models at equilibrium nuclear geometry and their comparison with the previously published results.}
		\label{tab:targcompar}
		\begin{center}
		  \begin{ruledtabular}
		    \begin{tabular}{@{}dddd}
					& \multicolumn{1}{l}{Model 1} & \multicolumn{1}{l}{Model 2} & \multicolumn{1}{l}{Bibliographical data}\\
					\hline
                                        \multicolumn{1}{l}{GTO basis}&\multicolumn{1}{l}{6-311G*}&\multicolumn{1}{l}{cc-pVDZ}&\\
					\multicolumn{1}{l}{\# of active electrons / \# of VOs}&8/5&4/11&\\
					\hline
					\multicolumn{1}{l}{Ground state energy (a.u.)}&-795.787343&-795.627044&\\
					\multicolumn{1}{l}{Vertical excitation energy $^1E$ (eV)}&9.545&9.057&7.7\pm0.1 \text{ \cite{lucena-ex, mckoy-spec}}\\
					\multicolumn{1}{l}{Vertical excitation energy $^3E$ (eV)}&8.476&7.897&\\
					\multicolumn{1}{l}{Dipole moment (a.u.)}&0.378&0.410&0.197 \text{ \cite{crc-book}}\\
					\multicolumn{1}{l}{Static dipole polarizability (a.u.)}&1.32&0.158&38.6 \text{ \cite{crc-book}}\\
		    \end{tabular}
		  \end{ruledtabular}
		\end{center}
	\end{table}
	\begin{table}[htb]
		\caption{Vibrational frequencies calculated using different CAS CI models and their comparison with the previously published results.}
		\label{tab:freqcompar}
		\begin{center}
			\begin{ruledtabular}
				\begin{tabular}{@{}dddd}
					& \multicolumn{1}{l}{Model 1} & \multicolumn{1}{l}{Model 2} & \multicolumn{1}{l}{Bibliographical data}\\
					\hline
					\multicolumn{1}{l}{$\omega_2$ (cm$^{-1}$)}&854.42&956.48&775.12 \text{ \cite{scanlon-cf3cl}}\\
					\multicolumn{1}{l}{$\omega_3$ (cm$^{-1}$)}&433.08&534.13&463.33 \text{ \cite{scanlon-cf3cl}}\\
					\multicolumn{1}{l}{\cft fragment $(\omega_2)$ (cm$^{-1}$)}&755.53&745&701\text{ \cite{suto-cf3}}\\
				\end{tabular}
			\end{ruledtabular}
		\end{center}
	\end{table}
	As it can be seen there, this target model (denoted as Model 1 there and elsewhere in 
the text) gives a reasonably good representation of the ground state dipole moment, as it is relatively small, although our calculation gives us a larger value than found in the experiment. The vertical excitation energy calculated here is almost 2\thinspace eV higher than the experimental value~\cite{mckoy-spec}. It suggests that the representation of the target excited states is limited. According to our knowledge, there is no experimental or advanced theoretical calculation of the lowest excited state $^3E$ available. Our target model gives its energy 8.453\thinspace eV above the ground state. In our scattering calculations, we are interested in scattering energies below 4\thinspace eV, where all the electronically excited channels are closed. We expect (and our test calculations described below suggest it) that inclusion of the low excited target states in the close-coupling (CC) expansion leads to a small correction of the resonance position and width only.
	
	In order to estimate how well are the polarization effects represented in our scattering calculations, we evaluated the static dipole polarizability of the target ground state. We used the sum-over-states formula with the set of target states included in the scattering calculation. We found that the lowest six target states represent a considerable contribution. Adding more target states did not increase the value. As can be seen in \tabab~\ref{tab:targcompar}, our calculated value is considerably smaller than the experimental results. 
	%Since the higher excited states don't contribute to the polarizability, this discrepancy can be possibly attributed to the primary GTO basis set which does not contain GTOs diffuse enough to describe the polarization potential.
	Since the higher excited states do not have significant contribution to the polarizability, this discrepancy can be possibly attributed to low number of active electrons and orbitals used in our target model. Good representation of the polarizability would require presence of more diffuse MOs which are too high in energy to be included in the present target calculations.
	However, the previously published one-dimensional calculations of the nuclear dynamics for \cfcl~\cite{wilde} show that the polarization effects have a minor effect on the results of the LCP calculations.
	
	Target properties discussed above are all calculated at the equilibrium nuclear geometry. However, for our calculations of the nuclear dynamics it is important to know how well can our target model reproduce the harmonic vibrational frequencies of the neutral molecule. Comparison of our calculated values with the previously published result is showed in \tabab~\ref{tab:freqcompar}. Our target model 1 gives the values with 10\% accuracy when compared with the experimental results. This small difference can be possibly attributed to the neglect of other vibrational degrees of freedom in our model.

	We can conclude that our CAS CI model 1 represents the neutral target 
sufficiently enough to be used in our \abini \rmat calculation. \rmat
calculations performed 
at the equilibrium nuclear geometry can be used to obtain the vertical attachment 
energy. Correct value of this quantity is essential for our calculations of the 
resonant nuclear dynamics. However, the CAS CI model 1 of the target leads to 
the value which is 1.3\thinspace eV above the experimental value determined 
by~\citet{aflatooni}. Additional testing calculations show that it is due to 
insufficient treatment of the electron correlation in the CC expansion of the 
$N+1$-electron wave function. Since further increase of the number of active 
electrons or VOs included in the target model leads to intractable 
diagonalization of the $N+1$-electron Hamiltonian, we decided to model the 
electron correlation by modification of the primary GTO basis set, where we 
modified the gaussian exponent corresponding to the p-orbital of the chlorine 
atom in order to get the energy of the lowest unoccupied molecular orbital (LUMO) closer to the experimental value of 
the vertical attachment energy. We expect that this modified LUMO helps to 
represent the discrete component of the resonant wave function of the temporal 
anionic complex better than the linear combination of VOs obtained in model 1. 
This idea is similar to the method developed in Ref.~\cite{simons-orbmodif} used to calculate the position and width of the resonance. The modified HF orbitals were calculated using the cc-pVDZ GTO basis set~\cite{dunning-cc}. To approach the correct value of vertical attachment energy we changed the exponent of the uncontracted p-orbital of chlorine from 0.162 to 0.094. In addition to this modification of the exponent, we also restricted the CAS CI model to four active electrons from the highest occupied molecular orbital (HOMO) and HOMO-1 and allowed their excitations into 11 lowest VOs. Using the notation of the $C_s$ point group, this CAS model can be expressed as follows:
	\begin{equation}
	  (1\mathrm{a}'\dots 16\mathrm{a}')^{32}(1\mathrm{a}''\dots7\mathrm{a}'')^{14}(17\mathrm{a}'\dots25\mathrm{a}',8\mathrm{a}''\dots11\mathrm{a}'')^4
	\end{equation}
	or
	\begin{equation}
	  (1\mathrm{a}'\dots 16\mathrm{a}')^{32}(1\mathrm{a}''\dots7\mathrm{a}'')^{14}(17\mathrm{a}'\dots24\mathrm{a}',8\mathrm{a}''\dots12\mathrm{a}'')^4,
	\end{equation}
	as the ordering of VOs changes with nuclear geometry. Therefore, in this model we extend the space of VOs and reduce the number of active electrons as compared with model 1. Properties of the target calculated at the equilibrium nuclear geometry are summarized in \tabab~\ref{tab:targcompar} (we refer to this model as to Model 2 in the table and elsewhere in the text below). It shows that although the ground state energy is higher than the value calculated using model 1, the vertical excitation energy into the $^1E$ state is lower and closer to the published reference value. This CAS model of the target also gives lower vertical excitation energy to the $^3E$ state than model 1. \tabab~\ref{tab:targcompar} also shows that this modified model gives larger value of the dipole moment than model 1, however its value is still low enough to not to introduce a significant error in the position and width of the resonance. We also calculated the static dipole polarizability at equilibrium nuclear geometry using this model and found that its value is significantly smaller than that given by experiment. This can be due to the limited representation of the polarization effects by the primary GTO basis set. Although both our CAS models of the target have their limitations, model 2 using the modified primary GTO basis represents correctly the vertical attachment energy, the quantity which is essential for the DEA calculations.
	
	In addition to the CAS model of the target we need a representation of the \cft fragment in our LCP calculations in order to extrapolate the potential energy surfaces properly. In order to keep the model of the fragment consistent with the target model 2, we treated the \cft radical at the SCF level as all the excitations of the active electrons in model 2 of the \cfcl target describe the C--Cl chemical bond. Therefore, any CI excitation model of the fragment would introduce a correlation which is not explicitly included in the target model. In order to check the quality of this representation we calculated the harmonic frequency of the umbrella mode. Its comparison with previously published value is given in \tabab\ref{tab:freqcompar}. It shows that our calculated harmonic frequency is slightly higher than the experimental value, but they are in a good agreement to confirm that our model is a suitable choice for representation of the \cft radical.

	\subsection{Scattering model}
	The \rmat calculations were performed using a sphere with radius 
$r_\Omega$=15\bohr. Corresponding continuum basis set was represented by 
single-center uncontracted GTOs with exponents optimized by the program GTOBAS~\cite{contibas}. 
Partial waves up to $l=3$ (9s, 7p, 7d, 7f) were used. The deletion threshold in the orthogonalization procedure for the continuum orbitals~\cite{tennyson-rev} was set to $\delta_{\text{thr}}=9\times 10^{-6}$. This value was found by performing calculations in the static exchange approximation. It gives a stable representation of the scattering continuum and does not show any problems related to linear dependence of the continuum orbitals.
	
	In order to obtain the position and width of the fixed-nuclei resonance 
as a function of the nuclear geometry we performed a scattering calculation of 
the eigenphases and fitted them using the Breit-Wigner formula with the energy 
dependent background~\cite{lane-thomas} as discussed above. We tested both 
CAS CI models of the neutral target discussed above. First, the lowest 16 CI 
target states calculated using Model 1 were used in the CC expansion and the
\rmat calculation at equilibrium nuclear geometry. This calculation gives 
converged results with respect to the number of target states included. 

\begin{figure}
		\includegraphics[scale=0.93]{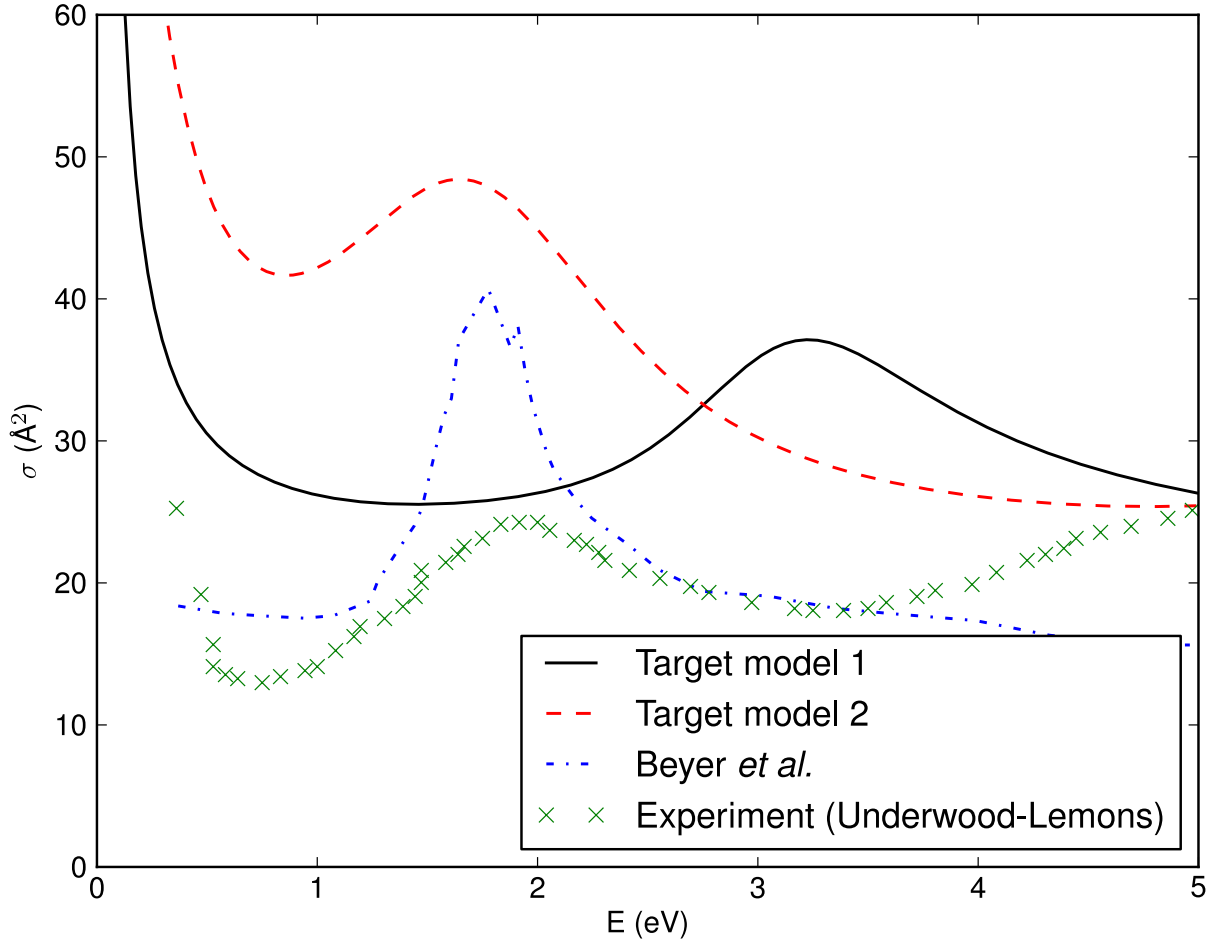}
		\caption{\label{fig:equics} Cross section for the elastic electron scattering off \cfcl ($^2A_1$ symmetry) calculated in the fixed-nuclei approximation at equilibrium nuclear geometry. Calculations employing target model 1 is compared with results calculated using model 2, with experimental results \cite{Und94} and with other \rmat calculation \cite{beyer-nestmann}.}
	\end{figure}

The corresponding cross section for elastic $^2A_1$ scattering is presented 
in \figab\ref{fig:equics} where we also compare it with calculations of~\citet{beyer-nestmann}.
	This curve (designated as Target model 1) shows the peak at energy 
around 3.3\thinspace eV that 
is significantly above the experimental value due to experimental work by 
\citet{Und94} and the theoretical calculation of \citet{beyer-nestmann}. 
Since a further increase of the number of target states included in the CC 
expansion did not lead to any considerable shift of the peak towards lower 
energies, it suggests that this artificially high position of the resonance 
is not due to incorrect representation of the polarization 
effects~\cite{rescigno-polar}, but rather due to an incompleteness of 
the target model. Therefore, we performed another set of scattering 
calculations using target CI model 2 described above. Since this model was 
constructed to reproduce the correct vertical attachment energy, it also gives the correct position of the resonance peak in the \rmat calculation carried 
at the equilibrium nuclear geometry (as plotted in \figab\ref{fig:equics}). 
The rapid 
increase of the cross section at lower energies is due to the long-range 
dipole potential included in the outer region. 
For a molecule with a nonzero permanent dipole moment and fixed orientation
the elastic cross section diverges at zero energy \cite{Mit65, Fab76},
the feature which is observed in our cross section, but not in the calculation
of~\citet{beyer-nestmann}, apparently because
the dipole effects
were not included completely there. On the other hand, since our dipole moment
at the equilibrium internuclear separation is too big (see Table I), it is
evident that our elastic cross sections are strongly overestimated at low 
energies. 

There were 12 target states 
included in our CC expansion (three lowest in singlet and triplet state of 
$A'$ and $A''$ symmetries). 
The \rmat was propagated in the dipole potential given by the target CAS CI 
model (as discussed above) and in the potential given by the dipole and 
quadrupole coupling of different scattering channels~\cite{tennyson-rev}.
On the other hand, the \rmat calculation of~\citet{beyer-nestmann} was performed at the level of static exchange with 
polarization which treats the electron correlation in a different way than 
our CAS CI model. This can be a partial reason of the quantitative difference 
between the two calculations at higher energies.

\figab\ref{fig:equics} also presents the 
experimental measurements of the total elastic cross sections by~\citet{Und94}. However, the problem with this and
two other \cite{mann-cf3cl,Jon86} measurements for CF$_3$Cl is that it is not 
quite clear what is measured there. The total elastic cross section for a polar
symmetric top is divergent even if rotations are included \cite{Cra67}. Only 
the inversion
splitting, which is extremely small for CF$_3$Cl, makes the elastic cross 
section finite \cite{Gal07}. This means that the scattering amplitude
at small scattering angle $\theta$ behaves like $1/\theta$, and only at an
extremely small angle $\theta_{inv}$ it becomes 
finite. In the experiment of~\citet{Und94} 
the elastic cross section is determined
from the transmitted current under the assumption that transmitted electrons are 
not scattered, therefore what is measured in fact is the cross section integrated
from a small angle $\theta_{\min}$ to $180^{\circ}$, where $\theta_{\min}$ 
is determined by the geometry of the experimental apparatus, and $\theta_{\min}$ is, most likely, significantly greater than $\theta_{inv}$. In the 
experiment of \citet{mann-cf3cl} the total cross section is 
obtained by
extrapolating the measured differential cross section to $\theta=0$. This
procedure also gives an underestimated total cross section since the actual differential cross section behaves as $1/\theta^2$ at small angles, if $\theta>\theta_{inv}$. We think, therefore that comparison between the theory and the experiment for the total (integrated) cross
section is meaningless unless the angle $\theta_{\min}$ can be found from
the experimental geometry. The purpose of plotting the experimental 
cross section in \figab\ref{fig:equics} is to demonstrate that our calculated 
resonance contribution is consistent with experimental results.

	At larger C--Cl separations, where the resonance turns into a bound state, the corresponding anionic bound state energies were calculated by diagonalizing the $N+1$ electron Hamiltonian constructed using the same target and scattering CAS CI model as was used for calculation of the resonance, just the integrals involving continuum GTOs were not restricted to the inner region of the sphere~\cite{tennyson-rev}. This allows a good representation of the diffuse character of the anionic bound state with energy close to the autodetachment limit.
	
	The position and width of the fixed-nuclei resonance as a function of 
the nuclear geometry were obtained by fitting the eigenphases at energies 
close to the resonance to the Breit-Wigner formula with energy-dependent 
background performed at each nuclear geometry of our interest. The 
energy dependence of the background is predominately determined by the dipole 
moment of the neutral target. The corresponding value was obtained from the 
calculation of the target properties for every nuclear geometry and used as 
a parameter in the fitting procedure for this geometry. Fitting of the 
eigenphases calculated at equilibrium geometry using target model 2 gives a 
resonance position 1.954\thinspace eV, which is in a good correspondence with experimental value 1.83\thinspace eV due to \citet{aflatooni}.

	\section{Calculations of the nuclear dynamics}
	\subsection{Dissociative electron attachment}
\begin{figure}
		\includegraphics[scale=0.9]{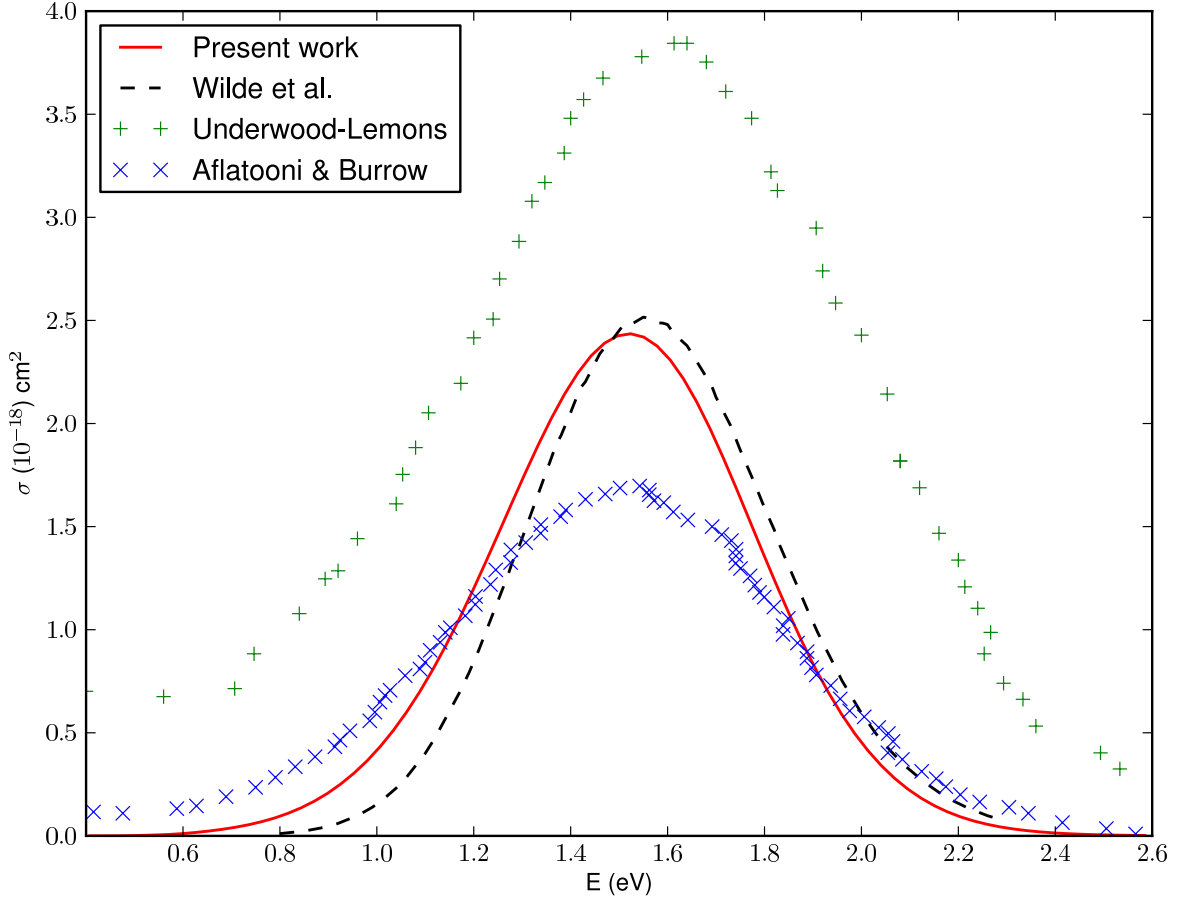}
		\caption{\label{fig:diffde}Total DEA cross section calculated using the model discussed in the present work is compared with experimental results of \citet{lemons}, \citet{aflatooni} and with previously published one-dimensional non-local calculations due to \citet{wilde}.}
	\end{figure}

	The total DEA cross section calculated using our two-dimensional LCP 
approach is plotted in \figab\ref{fig:diffde}. This graph shows its 	
	comparison with previously published one-dimensional non-local 
calculations~\cite{wilde}. Our previously published two-dimensional LCP 
calculations~\cite{tarana-cf3cl} give significantly larger magnitude of the 
total DEA cross section than experimental works~\cite{aflatooni,lemons} and 
one-dimensional non-local semiempirical calculation~\cite{wilde}. This problem 
was attributed to incorrect dependence of the width function (which was 
arbitrarily extended to two dimensions) on the coordinate $r$. As can be seen 
in \figab\ref{fig:diffde}, results of our present two-dimensional LCP 
calculations using the complex PES constructed from the \rmat results are in 
very good agreement with the non-local calculations as well as with 
experimental results due to~\citet{aflatooni}. Although our total cross 
section is smaller than the experimental results of~\citet{lemons}, it is closer to the measurements of~\citet{{aflatooni}}, and the position of the peak is in very good correspondence with this experimental work. All this suggests that the two-dimensional PES constructed from the results of the \rmat calculations has more correct dependence on both reaction coordinates than the previously published model~\cite{tarana-cf3cl}. It results in a correct shift of the peak in the total DEA cross section with respect to the VAE as well as in the correct magnitude of the cross section. 
\begin{figure}
  \includegraphics[scale=0.8]{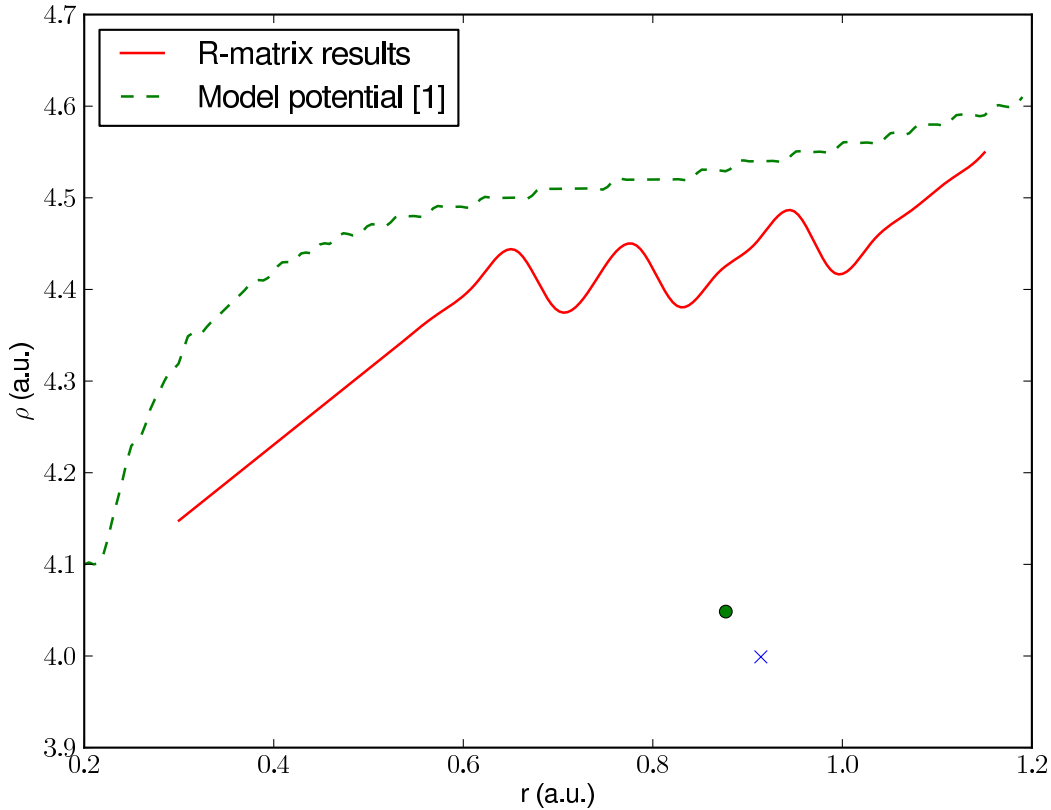}
  \caption{The crossing seam between the neutral and anionic two-dimensional PES obtained from the present \rmat calculations and its comparison with model potential used in~\cite{tarana-cf3cl}. The cross (circle) denotes the equilibrium geometry of the neutral PES obtained in the present work (used in~\cite{tarana-cf3cl}).}
  \label{fig:seam}
\end{figure}
One additional note regarding the complex PES should be made here. The DEA cross section is related to the position of the crossing seam between the neutral and anionic PES. Its shape and position with respect to the minimum of the neutral PES determine where the anionic system becomes stable. Subsequently, it has an influence on the survival probability. The crossing seams obtained from the present \rmat calculations and from the model potential~\cite{tarana-cf3cl} are compared in \figab\ref{fig:seam}.
The relative position of the crossing seam is similar in both models. This supports our argument that the difference in magnitude of the DEA cross section between our present model and the results published in~\cite{tarana-cf3cl} is mainly due to the different width function $\Gamma(\rho,r)$, rather than due to the substantial difference in the crossing seam between these two models. The oscillatory structure in the crossing seam obtained from the present R-matrix calculations is mainly an artifact of too low density of the grid of nuclear geometries used to calculate the PES. In addition, the fact that the complex energies in the region, where the anion is metastable are obtained in different way than the eneries of the bound anion, also raises the numerical issues with exact determination of the crossing seam and partially also contibutes to the oscillatory structure present in \figab\ref{fig:seam}.

Comparisons of local two-dimensional and one-dimensional results with nonlocal one-dimensional results are given in Refs.~\cite{tarana-cf3cl,fabrikant-conf}. The agreement between the local and non-local results is very good because the resonance occurs at a relatively large energy, and the long-range interaction plays a minor role.
	
\begin{figure}
		\includegraphics[scale=0.93]{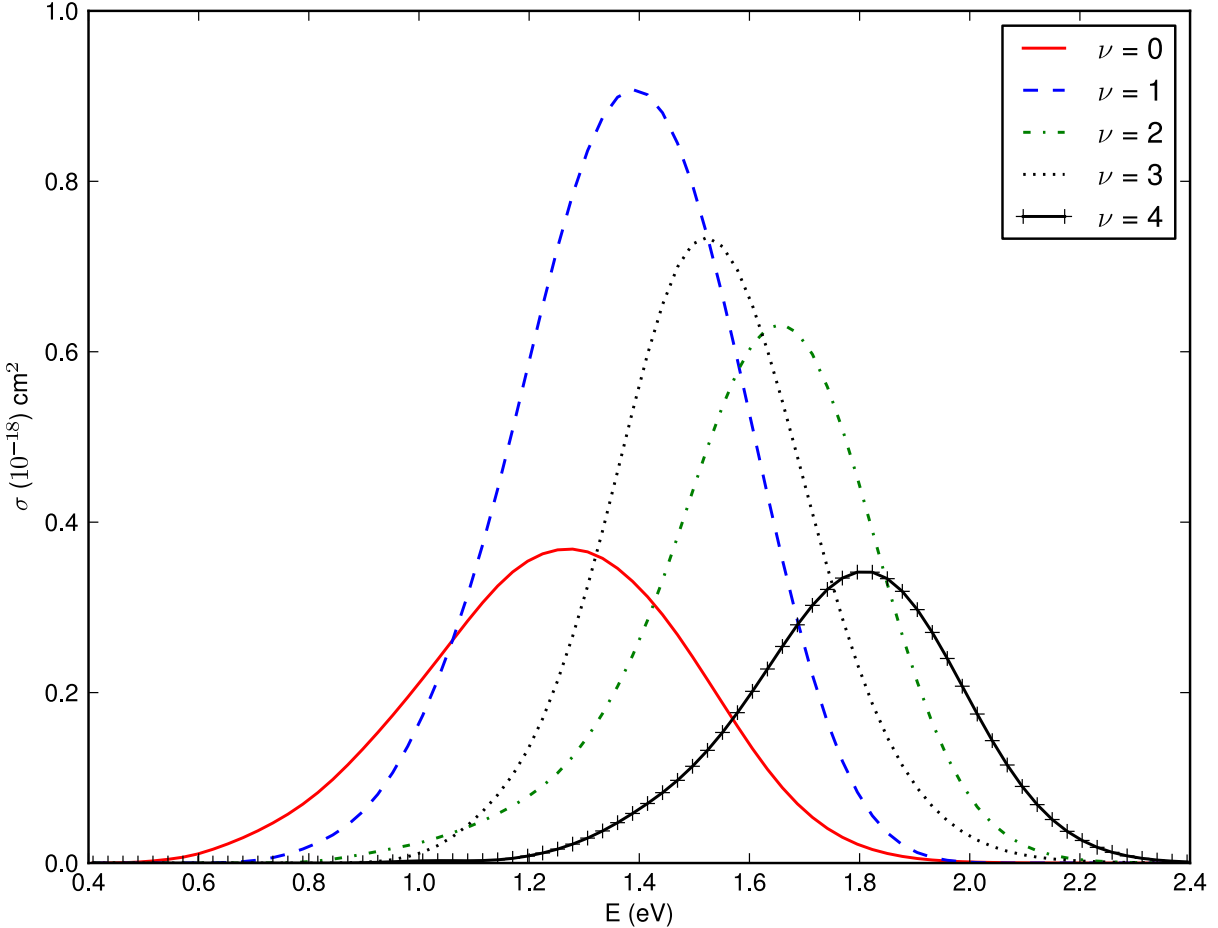}
		\caption{\label{fig:chandea}DEA cross sections for different final vibrational states of the fragment \cft calculated using the complex PES constructed from our \rmat results.}
	\end{figure}

\begin{figure}
		\includegraphics[scale=0.93]{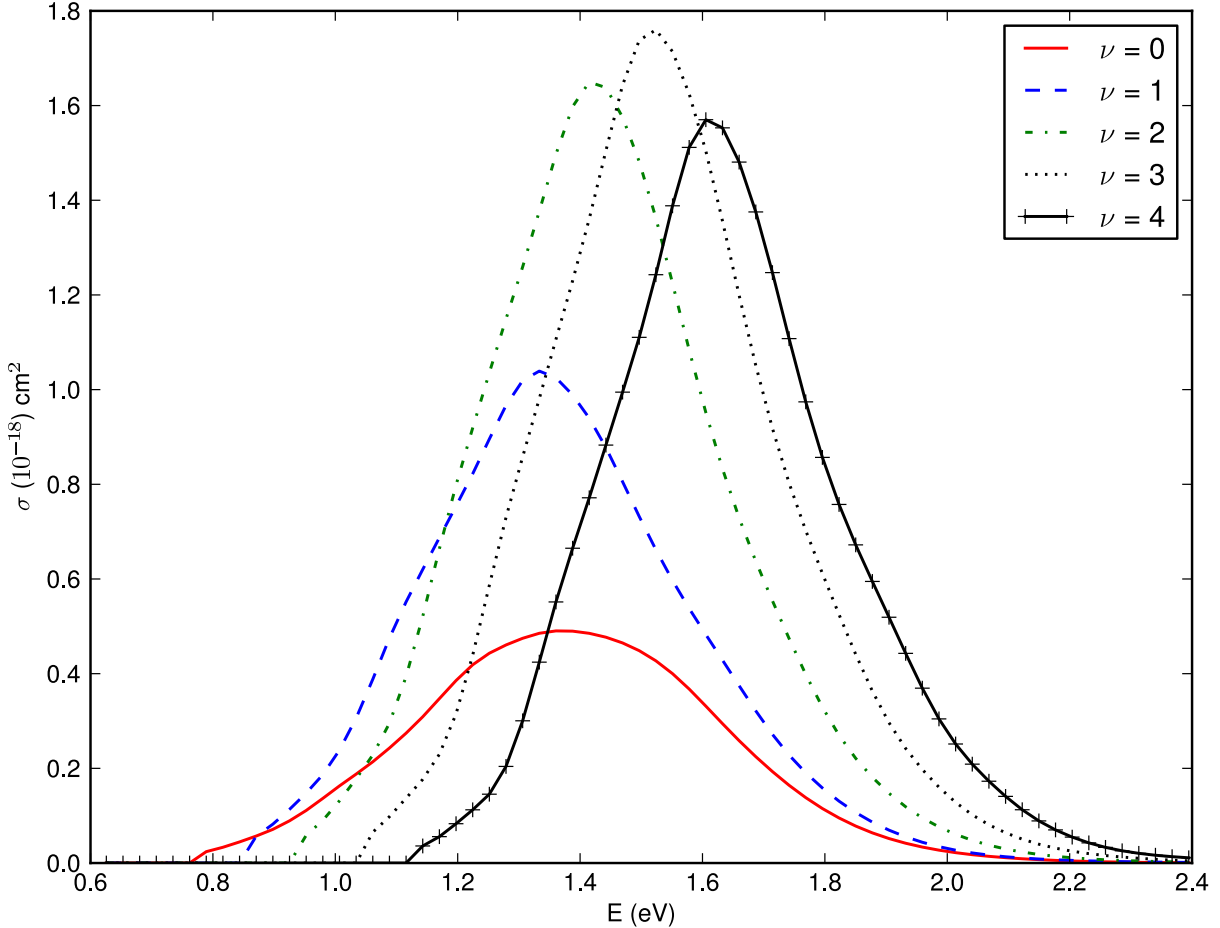}
		\caption{\label{fig:oldchandea}DEA cross sections for different final vibrational states of the fragment \cft calculated using the model complex PES published previously~\cite{tarana-cf3cl} after the correction of the asymptotic behavior of channel potentials.}
	\end{figure}

	\figab\ref{fig:chandea} shows the distribution of different final 
vibrational states of the \cft fragment calculated using the complex PES 
constructed from the \rmat results. In \figab\ref{fig:oldchandea} we show the 
distribution calculated 
using 
the PES obtained in Ref.~\cite{tarana-cf3cl} but corrected in such a way
that PES in the intermediate region smoothly turns into the PES in the
asymptotic region avoiding the mismatch discussed in Sec.~\ref{sec:theo}. 
 Both graphs show that the low vibrationally excited states of the \cft 
fragment will be more populated than the ground state and in both figures the 
positions of the peaks rise with increasing quantum number $\nu$. However, 
each calculation predicts highest population for a different excited state. While our complex PES constructed from the \rmat results gives the highest peak for $\nu = 1$, the PES described in~\cite{tarana-cf3cl} leads to the highest peak for vibrational state $\nu = 3$, as can be seen in \figab\ref{fig:oldchandea}. Total DEA cross section is mainly determined by the complex PES in the region where the anionic system is not bound and does not strongly depend on the behavior of the PES in the region where the negative ion is stable. On the other hand, the distribution of vibrational states of the fragment is strongly influenced by the final-state interaction in the region where the anion is stable. In both calculations shown in \figab\ref{fig:chandea} and \figab\ref{fig:oldchandea}, this region of the PES was partially modeled (as discussed above and in~\cite{tarana-cf3cl}) to achieve correct asymptotic behavior of the potential. Although our PES constructed from the \rmat results is free of several limitations of the surface described in~\cite{tarana-cf3cl}, the extrapolation of the bound-state anionic surface in both cases makes it difficult to decide, how quantitatively reliable this distribution is in both calculations.

	\subsection{Vibrational excitation}
\begin{figure}
		\includegraphics[scale=0.93]{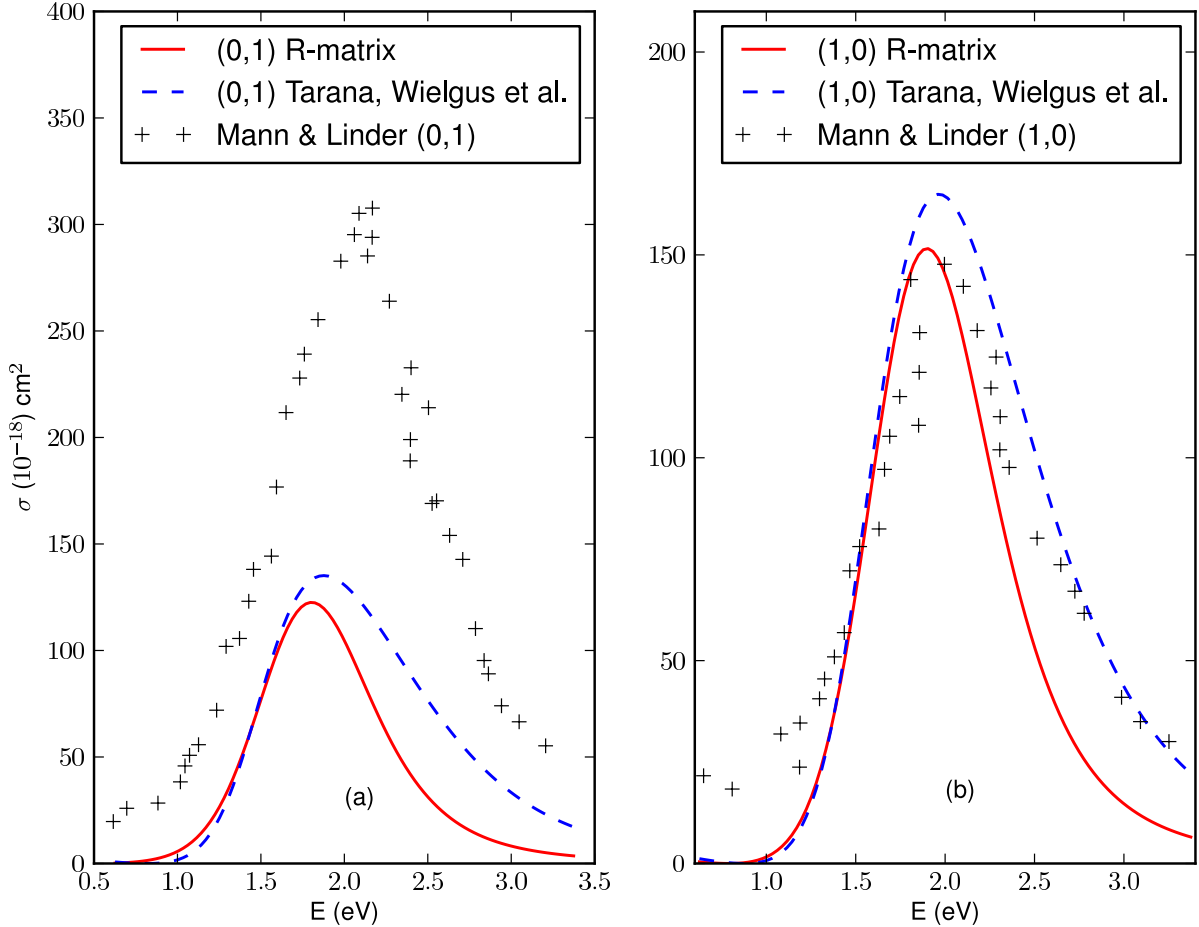}
		\caption{\label{fig:vecompar}VE cross sections from the ground vibrational state to the lowest excited state of the C-Cl stretching mode (a) and the lowest excited state of the umbrella mode (b). Results calculated using the \rmat complex PES are compared with cross sections obtained from the previously published model~\cite{tarana-cf3cl} and experimental results due to \citet{mann-cf3cl}.}
	\end{figure}

	Results of the resonant VE calculations using the PES based on 
the \rmat results are plotted in \figab\ref{fig:vecompar}. This figure shows 
the cross section for VE of the target from the ground state to the lowest 
excited state of the C--Cl stretching mode (denoted as $(0,1)$ in the figure) 
and the lowest excited state of the C--F deformation mode (denoted as $(1,0)$ 
in the figure). These graphs show a very good agreement with our previous 
calculations~\cite{tarana-cf3cl} for both vibrational modes. 
However, our calculations agree with experimental results by \citet{mann-cf3cl} 
for the umbrella mode, while both our models lead to approximately three times 
smaller magnitude of the cross section for the C--Cl stretching mode excitation 
than that measured by \citet{mann-cf3cl}.
	
	The previously published one-dimensional non-local calculation of 
VE \cite{wilde} takes into account the C-Cl stretching vibrational mode only and the width function $\Gamma(R)$ was adjusted to give cross sections corresponding to experimental results for this stretching vibrational modes excitation due to \citet{mann-cf3cl}.
%Therefore, our cross sections calculated using both two-dimensional models are smaller the results of the one-dimensional calculation. 
The low values of the present cross sections can be partially explained by introduction of additional channels of vibrational excitation in our two-dimensional calculations and lower flux towards the channel $(0,1)$.
	
\begin{figure}
	  \includegraphics[scale=0.9]{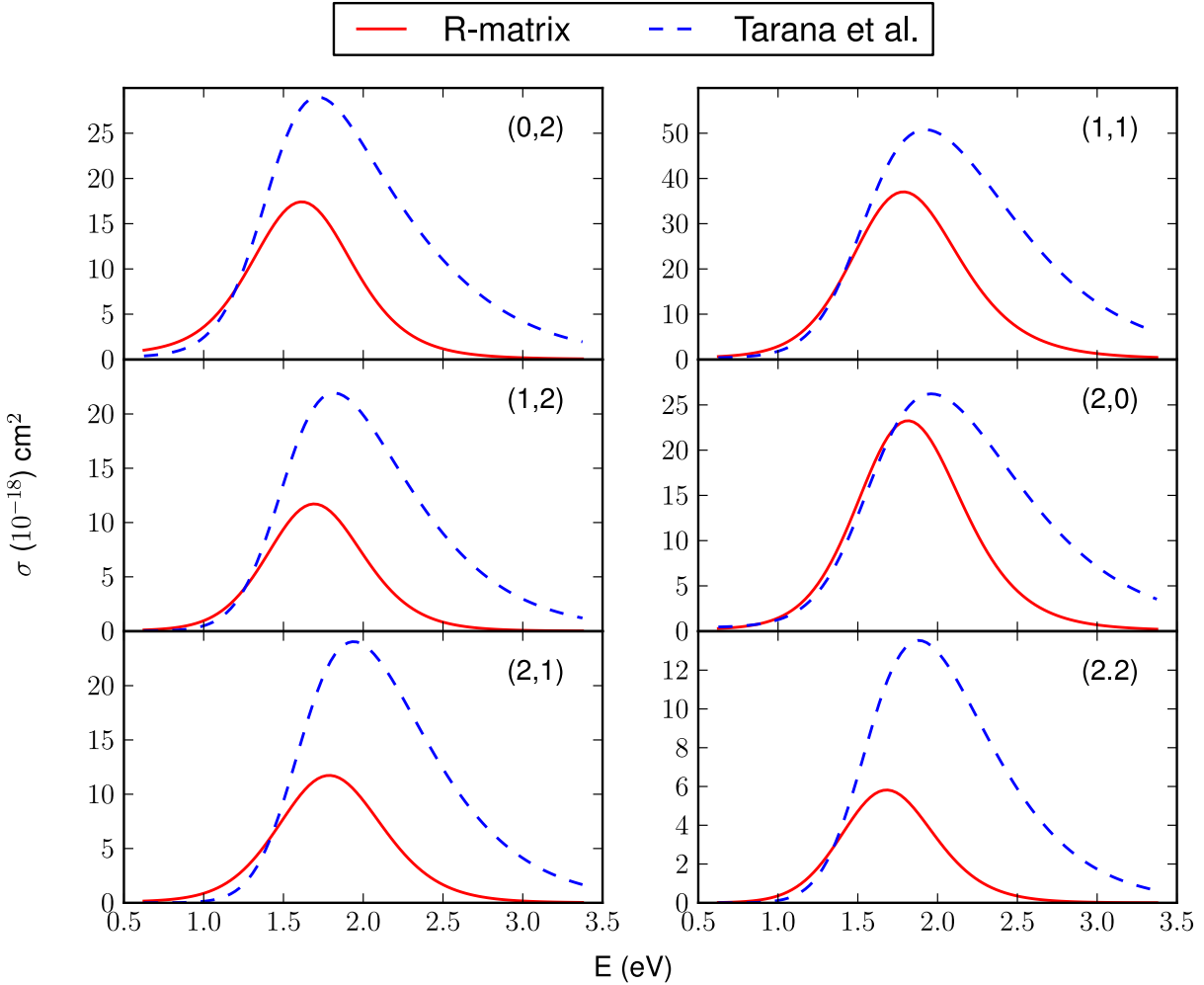}
	  \caption{\label{fig:hivecomp}VE cross sections for excitation from the target vibrational ground state to higher excited states. Results obtained from the complex PES constructed from the \rmat results (solid line) are compared with cross sections calculated using the PES described in~\cite{tarana-cf3cl} (dashed line).}
	\end{figure}

	Although \figab\ref{fig:vecompar} shows that the model constructed 
using the \rmat results leads to VE cross sections very similar to those 
calculated in Ref. \cite{tarana-cf3cl}, \figab\ref{fig:hivecomp} shows that 
these two models predict different results for higher final vibrational states. 
To our best knowledge, there are no experimental data for VE to these states 
to compare with our cross sections. The fact that the differences between our 
two models become more significant with increasing final vibrational state is 
understandable, since the target vibrational eigenfunctions become spatially more extended with increasing vibrational state $(\nu_2, \nu_3)$. This means that the different behavior of corresponding complex PES farther from equilibrium will have larger influence on results, as can be seen in \eqab\eqref{eq:lcpbas}.

	\section{Conclusion}
The \abini \rmat method allowed us to calculate the 
complex two-dimensional PES for \cfcl collisions. We used then DVR method
to obtain the solution of coupled stationary equations \eqref{eq:lcpbas}, and
DEA and VE cross sections for the $e-$\cfcl collision process. Our results 
for the total DEA cross section and the cross section for VE of the umbrella mode 
agree quite well with experiments, In addition we obtained the final-state 
vibrational distribution in the \cft fragment free of instabilities found
in our previous calculations. However our cross sections for VE of the C-Cl 
stretching mode are significantly lower than the experimental results of
\citet{mann-cf3cl} and the results of previous semiempirical
one-dimensional calculations \cite{wilde}. This is something one might expect 
because of the extra inelastic channels of excitation of umbrella mode that
leads to a redistribution of the flux. However, disagreement of the present
2-dimensional results with the experiment is puzzling and requires a further
investigation.

\begin{acknowledgments}
This work was supported in part by the Department
of Energy, Office of Science, the National Science Foundation under Grants
No. PHY-0652866, No. PHY-0969381, and by a Marie Curie International Incoming Fellowship
(FP7-PEOPLE-2009-IIF-252714). Support from the Czech Science Foundation (GA\v{C}R) by Grant No.
208/10/1281 and by Z\'{a}m\v{e}r MSM0021620860 of the Ministry of
Education, Youth and Sports of the Czech Republic is also gratefully
acknowledged.

\end{acknowledgments}

%	\bibliography{refs}

%

\end{document}